\documentclass[aps,prx,showpacs,amsmath,amssymb,amsfonts,lengthcheck,twocolumn,longbibliography,superscriptaddress]{revtex4-2}

\usepackage{graphicx,color}
\usepackage{amsthm}
\usepackage{amsfonts}
\usepackage{algorithmic}
\usepackage{enumerate}
\usepackage{latexsym}
\usepackage{amsmath}
\usepackage{amssymb}
\usepackage[normalem]{ulem} 
\usepackage[dvipsnames,svgnames,x11names]{xcolor}
\usepackage[colorlinks=true,citecolor=blue,linkcolor=blue,urlcolor=blue]{hyperref}
\usepackage[T1]{fontenc}
\usepackage[english]{babel}
\usepackage{diagbox}
\usepackage{amssymb}
\usepackage{comment}
\usepackage{cancel}

\newcommand{\Tr}        {\mathrm{Tr}}
\newcommand{\bra}[1]    {\langle #1|}
\newcommand{\ket}[1]    {| #1 \rangle}
\newcommand{\bk}[2]     {\langle #1 | #2 \rangle}

\newcommand{\cS}        {{\mathcal S}}

\newcommand{\cE}        {{\mathcal E}}

\newcommand\cF{{\mathcal F}}
\newcommand\hocom[1]{}

\newcommand{\ba}{\begin{eqnarray}}
	\newcommand{\ea}{\end{eqnarray}}
\newcommand{\bmath}{\begin{mathletters}}
	\newcommand{\emath}{\end{mathletters}}
\newcommand{\ban}{\begin{eqnarray*}}
	\newcommand{\ean}{\end{eqnarray*}}

\newcommand{\bla}{bla\\bla\\bla\\bla\\bla}
\newcommand{\PRA}{Phys. Rev. A }

\usepackage{fmtcount}

\newcommand{\mc}[1]{\mathcal{#1}}

\newcommand{\MI}{I(\mathcal{S}\!:\!\mathcal{F})}
\newcommand{\discord}{\mathcal{D}(\mathcal{S}\!:\!\check{\mathcal{F}})}
\newcommand{\holevo}{\chi(\mathcal{S}\!:\!\check{\mathcal{F}})}

\newcommand{\draftmode}{1}    
\newcommand{\notetoself}[1]{\ifnum \draftmode=1 {\color[rgb]{0,0,0.8} [#1]} \fi}  
\newcommand{\cuttext}[1]{\ifnum \draftmode=1 {\color[rgb]{0,0.5,0} [#1]} \fi}  
\newcommand{\warntext}[1]{\ifnum \draftmode=1 {\color[rgb]{0.9,0.6,0} #1} \else {#1} \color{black} \fi}
\newcommand{\aref}[1]{{Appendix~\hyperref[#1]{A}}}
\newcommand{\bref}[1]{{Appendix~\hyperref[#1]{B}}}
\newcommand{\dref}[1]{{Appendix~\hyperref[#1]{C}}}
\definecolor{darkbrown}{rgb}{0.4, 0.26, 0.13}

\begin{document}

\title{Observation of Quantum Darwinism and the Origin of Classicality with Superconducting Circuits}

\author{Zitian Zhu}\thanks{These authors contributed equally}
\affiliation{School of Physics, ZJU-Hangzhou Global Scientific and Technological Innovation Center, and Zhejiang Key Laboratory of Micro-nano Quantum Chips and Quantum Control, Zhejiang University, Hangzhou, China}

\author{Kiera Salice}\thanks{These authors contributed equally}
\affiliation{Department of Physics, University of Houston, Houston, TX 77204}

 \author{Akram Touil}\thanks{These authors contributed equally}
 \affiliation{Theoretical Division, Los Alamos National Laboratory, Los Alamos, New Mexico 87545}

\author{Zehang Bao}
\author{Zixuan Song}
\author{Pengfei Zhang}
\author{Hekang Li}
\author{Zhen Wang}
\author{Chao Song}

\author{Qiujiang Guo}
\email{qguo@zju.edu.cn}
\affiliation{School of Physics, ZJU-Hangzhou Global Scientific and Technological Innovation Center, and Zhejiang Key Laboratory of Micro-nano Quantum Chips and Quantum Control, Zhejiang University, Hangzhou, China}

\author{H. Wang}
\affiliation{School of Physics, ZJU-Hangzhou Global Scientific and Technological Innovation Center, and Zhejiang Key Laboratory of Micro-nano Quantum Chips and Quantum Control, Zhejiang University, Hangzhou, China}
\affiliation{State Key Laboratory for Extreme Photonics and Instrumentation, Zhejiang University, Hangzhou, China}
\author{Rubem Mondaini}
\email{rmondaini@uh.edu}
\affiliation{Department of Physics, University of Houston, Houston, TX 77204}
\affiliation{Texas Center for Superconductivity, University of Houston, Houston, Texas 77204, USA}

\begin{abstract} 
\textbf{The transition from quantum to classical behavior is a central question in modern physics. How can we rationalize everyday classical observations from an inherently quantum world? For instance, what makes two people, each absorbing an independent fraction of photons scattered from this screen or paper, agree on the observation of the text written here? Quantum Darwinism offers a compelling framework to explain this emergence of classicality by proposing that the environment redundantly encodes information about a quantum system, leading to the objective reality we perceive.
Here, by leveraging cutting-edge superconducting quantum circuits, we observe the highly structured branching quantum states that support classicality and the saturation of quantum mutual information, establishing a robust verification of the foundational framework of quantum Darwinism and the accompanying underlying geometric structure of quantum states. 
Additionally, we propose a particular class of observables that can be used as a separate quantifier for classicality, originating a computationally and experimentally inexpensive method to probe quantum-to-classical transitions. Our investigation delves into how the quantum effects are inaccessible to observers, allowing only classical properties to be detected. It experimentally demonstrates the physical framework through which everyday classical observations emerge from underlying quantum principles and paves the way to settling the measurement problem.}
\end{abstract}

\maketitle
Quantum mechanics dramatically upsets our intuitive understanding of nature, changing the long-held view that classical reality is an independent and objective state we merely observe. As we near the 100$^{\rm th}$-anniversary of quantum mechanics, the longstanding question of how the classical world emerges from the quantum realm remains one of the most profound challenges in modern physics. Parts of this puzzle are now more evident: It is clear that quantum systems cannot be fully understood in isolation; their interactions with the environment must often be considered, leading to the development of quantum decoherence theory~\cite{Zurek2003RMP,schlosshauer2007,joos2013decoherence}. By treating the universe as a collection of interacting quantum systems, one thus considers how the environment monitors certain observables of a system of interest. This monitoring destroys quantum coherences, causing the emergence of a preferred set of stable states~\cite{basis1,basis2}, dubbed `pointer states' (supplementary text section 1A). The process by which these survive is termed `einselection' -- short for environment-induced superselection~\cite{Zurek2000AP}. 

Decoherence is fundamental because it explains why quantum systems, despite their coherent nature, give rise to classical-like behavior. Namely, it ensures that quantum superpositions turn into classical joint probability distributions localized at specific outcomes, explaining why we never observe macroscopic superpositions in our daily lives. Still, it does not fully answer how classicality is perceived in a quantum universe. 
\begin{figure*}[t!]
    \begin{center}
    \includegraphics[width=0.89\textwidth]{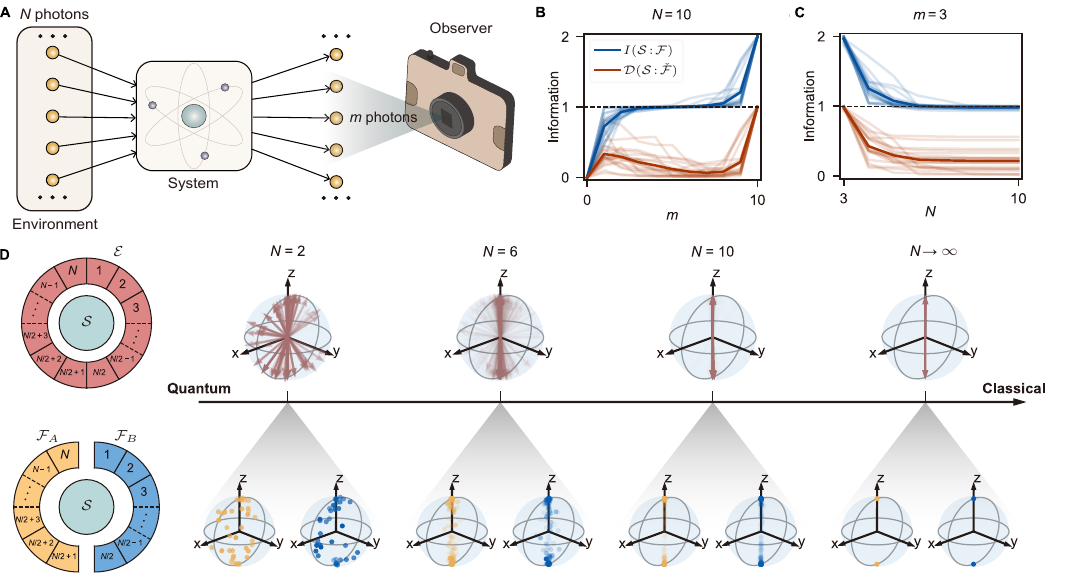}
    \end{center}
     \caption{\textbf{Quantum Darwinism and the emergent classical reality.}
    ({\bf A})
    An observer eavesdrops on the state of a system via a fragment of the environment. In the macroscopic world, we observe the world around us through information carried from $N$ photons interacting with systems of interest. An observer can only access a small fraction of them ($m$ photons) as evidence to describe the system's state. 
    ({\bf B})  Mutual information $\MI$ and quantum discord $\discord$ as a function of the fragment size $m$. 
    Conditional gates $\{U^{k}_{\oslash}\}$ -- see text -- are randomly sampled to simulate the interactions between the system and photons with relevant quantities numerically calculated over 20 runs; faint lines refer to each random realization, and the solid one is their average. The mutual information reaches $H_{\mathcal{S}}\simeq 1$, the system's entropy (here a single qubit) when a small fragment $\cF$ of the environment is observed. At this stage, the imprinted information on $\mathcal{F}$ is mostly classical because of the vanishing quantum discord. As the fragment size $m\to N$, $\discord$ suddenly rises, leading to increased mutual information beyond the classical plateau: It indicates the existence of quantum correlations between the system and the whole environment.
    ({\bf C}) $\MI$ and $\discord$ as a function of environment size $N$ with fragment size $m=3$. Only classical information of $\mathcal{S}$, $H_{\mathcal{S}}\simeq1$, can be redundantly recorded in $\cF$ at large $N$, which enables the objective existence in our daily life.
    ({\bf D}) A branching structure emerges as the quantum-to-classical transition takes place. In the top row, the whole environment $\cE$ is observed ($m=N$). With growing $N$, the system experiences increasing decoherence such that when $N\rightarrow\infty$, $|\psi_{\mathcal{SE}}\rangle$ is at the branching state, clustering around the pointer states $|0\rangle$ and $|1\rangle$ of $\cS$. In the bottom row, the environment is separated into two halves $\mathcal{F}_A$ and $\mathcal{F}_B$, independently measured by different observers. If the size of the environment is not sufficiently large, e.g., $N=2$ or $N=6$, observation results strongly depend on the type of interaction due to quantum effects, making them different for various observers. For $N\rightarrow\infty$, all results of different observers agree, which is what we infer in the classical world.}
    \label{fig:fig1}
\end{figure*}

For that, quantum Darwinism~\cite{Zurek2000AP,Ollivier2004PRL,Ollivier2005PRA,Zurek2009NP,zurek2022entropy,touil2022,Giorgi2015PRA,Balaneskovic2015EPJD,Balaneskovic2016EPJD,Knott2018PRL,Milazzo2019PRA,Campbell2019PRA,Ryan2020,Garcia2020PRR,Lorenzo2020PRR,davide2022,duruisseau2023pointer,Zurek_2025} expands upon decoherence, by asserting that the environment not only causes decoherence but also redundantly encodes classical information about the system's pointer states across its distinct fragments. This redundancy allows different observers to indirectly and independently access and confirm the classical state of the system without disturbing it. Decoherence, therefore, plays a dual role in quantum-to-classical transitions: suppressing quantum coherence while ensuring that classical information is robust and accessible in the environment, laying the foundation for the emergence of objective classical reality. 

More specifically, this can be observed as an example from daily life, as illustrated in Fig.~\ref{fig:fig1}A, where a camera represents an observer measuring a central system; it can also be equivalently interpreted as taking a picture of an object, say a tree. The camera captures photons scattered from the tree and indirectly learns about the tree's position through these photons. Now, consider a second observer taking a picture of the tree simultaneously. The photons absorbed by both cameras will differ, but the observers still agree upon the tree's position, as they have learned the same information. This information can be quantified by the mutual information, $\MI=H_{\mathcal{S}}+H_{\mathcal{F}}-H_{\mathcal{S F}}$, which is the total bipartite information of the system $\cal S$ and a fragment $\cal F$ of the environment $\cE$ (total photon bath). Here, $H_{{i}}=-\Tr\left[\rho_{{i}}\log_2\left(\rho_{{i}}\right)\right]$, is the von Neumann entropy of subsystem ${i}$. When all observers learn the same information, a plateau emerges in the mutual information as a function of the fraction of photons captured (Fig.~\ref{fig:fig1}B). In general settings, such classical reality only manifests in a large enough environment, i.e., $N\gg m$ (Fig.~\ref{fig:fig1}C), which is the typical scenario in the macroscopic world. Additionally, the information about purely quantum correlations, known as quantum discord~\cite{Ollivier2001PRL}, $\discord=\MI-\holevo$, tends to zero -- this is a precise definition of classicality in quantum Darwinism. Here, $\holevo=H_{\mathcal{S}}-{\rm min}_{\{M_k^{\mathcal{F}}\}}(H_{\mathcal{S} \mid \check{\mathcal{F}}})$, is the Holevo bound, which quantifies the maximum classical information one can obtain from an optimal quantum measurement chosen from the set of measurements $\{M_k^{\mathcal{F}}\}$ on $\mathcal{F}$, where $H_{\mathcal{S} \mid \check{\mathcal{F}}}$ is the conditional entropy.

The physical mechanism underlying the characterization of classicality can be uncovered by employing insights from geometric quantum mechanics~\cite{fabio1,fabio2,fabio3} alongside the aforementioned information-theoretic quantities. It has been formally shown~\cite{touil2022branching} that quantum states tend to cluster around specific classical configurations -- such states are formed from the pointer states that survive environmental monitoring and are referred to as branching states~\cite{blume2005simple,blume2006quantum}. In particular, a unique structure of states of the system and environment exists such that local quantum correlations are suppressed, and classical information is redundantly copied in the many information-bearing degrees of freedom of the environment; an example is given in Fig.~\ref{fig:fig1}D for the case with system-environment interactions mapped by random quantum unitaries as defined below.

Despite the recent theoretical advancements in delineating the origin of classicality, preliminary experimental results \cite{Ciampini2018PRA,Chen2019SB,Unden2019PRL} only show limited information-theoretical signatures of quantum Darwinism in special settings, such as encoding redundancy on specific GHZ initial states or realizing observation in a small number of environmental degrees of freedom. In particular, the geometric underpinnings of the global quantum wavefunction supporting quantum-to-classical transitions are still unexplored. In addition, from an operational perspective, a demonstration that connects the arguments of quantum Darwinism with the practical observing process is still lacking.  

In this Article, we present a comprehensive experimental demonstration of quantum Darwinism and study the self-organizing branching of quantum states through the lens of geometric quantum mechanics. Leveraging the tuning flexibility of our high-quality superconducting quantum qubits \cite{xu2023digital,Bao2024NC}, which feature energy relaxation time $T_1$ around $130~\mu$s,  and fidelities of single-qubit gate around 0.9997 and two-qubit CZ gate around 0.998 (see supplementary text section 2A), we observe that the formation of classical reality accompanies a clustering around the pointer states of the system's wavefunction, and further show the encoded classical information of system can be exactly decoded from environment fragments. In particular, this clustering is a consequence of decoherence, eventually resulting in zero quantum discord. Building upon this insight, we propose a novel approach for quantifying quantum Darwinism through suitably chosen local observables, facilitating its verification, offering further evidence for the theory, and leading to potential new applications.
\begin{figure*}[t!]
    \begin{center}
    \includegraphics[width=1\textwidth]{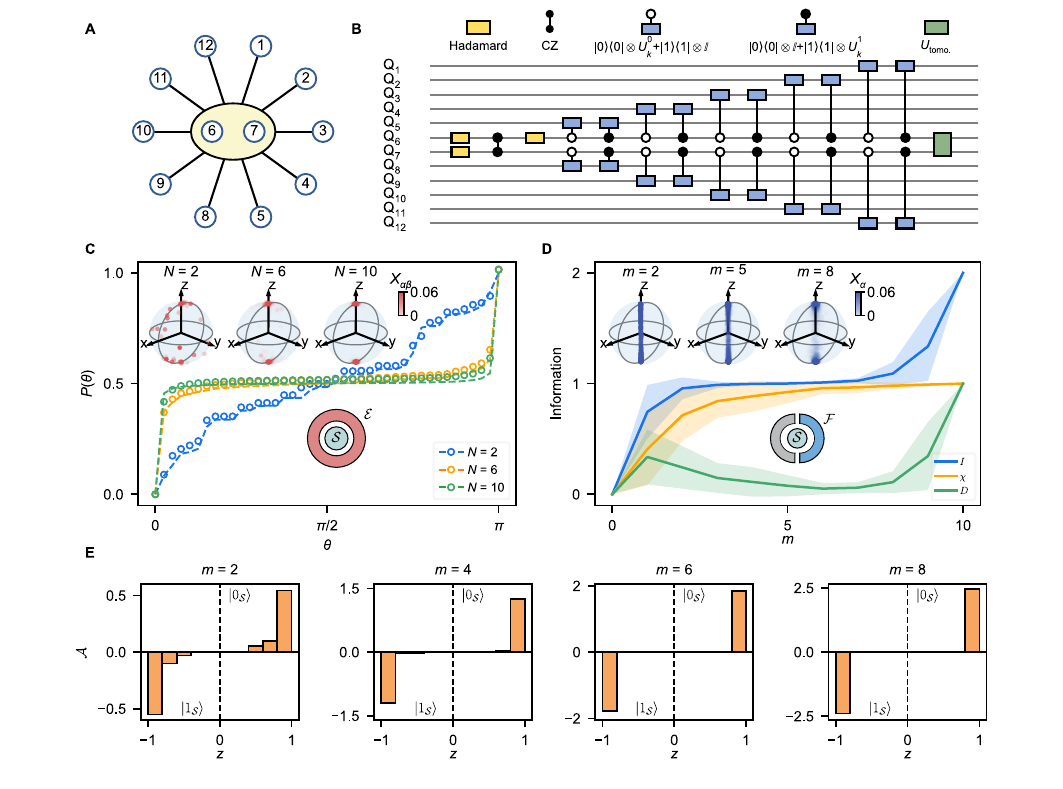}
    \end{center}
   \caption{\textbf{Quantum-to-classical transition: emergence of branching structures.}
    ({\bf A}) A diagram detailing the system-environment interaction topology. We choose two qubits, Q${}_6$ and Q${}_7$, as the system and the remaining ten as the environment. Black lines represent the interaction between the system and the environment. 
    ({\bf B}) Quantum circuit used to simulate the interaction. Similar to Fig.~\ref{fig:fig1}, $\theta_k^j$ and $\phi_k^j$ in $U_k^j$ are randomly sampled from the uniform distribution $\left[(j-0.5)\pi, (j+0.5)\pi\right)$ and $\left[-\pi,\pi\right)$, respectively.
    ({\bf C}) Experimental results of the integrated probability $P(\theta)$ and the distribution of geometric state $\mu_{\cS}$ on the Bloch sphere (inset). Blue, orange, and green points (dashed lines) represent experimental (noisy simulation, see supplementary text section 2F) results with increasing environment size $N$. Red dots in the Bloch spheres depict the experimentally reconstructed $\{X_{\alpha \beta}, \rho_{\cS}^{\alpha\beta}\}$.
    ({\bf D}) Measured $\{X_{\alpha}, \rho^{\alpha}_{\mathcal{S}}\}$ for fragment size $m=2, 5, 8$ (inset) and numerical results of the mutual information $\MI$, Holevo bound $\holevo$, and quantum discord $\discord$ as a function of $m$; the environment size is fixed with $N=10$. Solid lines are the average values over 10 random realizations where shadows indicate the standard deviations for each quantity. The blue dots in the Bloch spheres (inset) represent the experimentally reconstructed $\{X_{\alpha}, \rho_{\mathcal{S}}^{\alpha}\}$.
    ({\bf E}) Histograms of distributions of $\mathcal{A}(z)$ along $z$-axis of the Bloch sphere for fragment size $m=2,4,6,8$. The sign of $\mathcal{A}(z)$ builds up a one-to-one correspondence with the pointer states $\{\ket{0_{\mathcal{S}}}, \ket{1_{\mathcal{S}}}\}$.}
    \label{fig:fig2}
\end{figure*}

\vspace{.5cm}
\noindent\textbf{Branching states and quantum Darwinism}
\vspace{.2cm}
\begin{figure*}[t!]
    \begin{center}
    \includegraphics[width=0.96\textwidth]{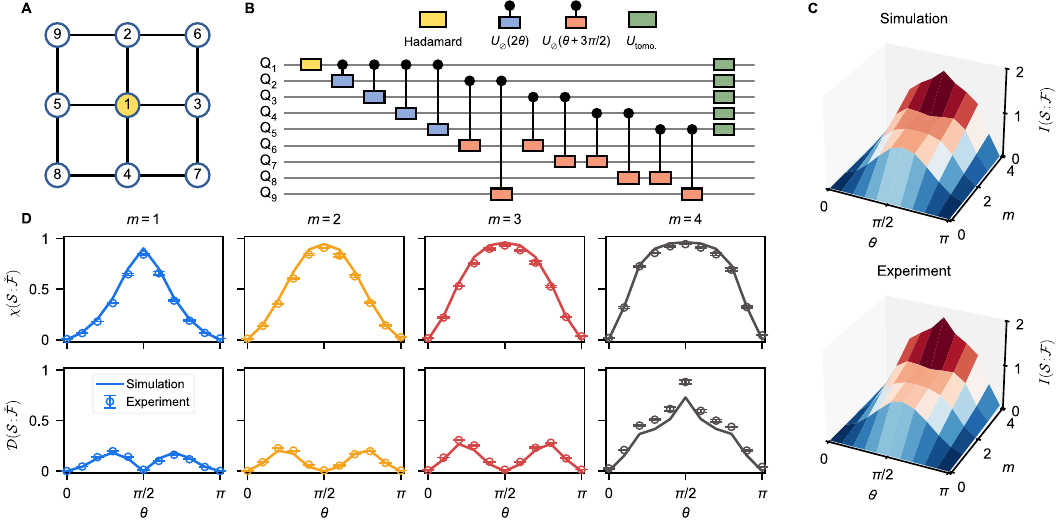}
    \end{center}
    \caption{\textbf{Robustness of the predictions of quantum Darwinism.}
    ({\bf A}) The 9-qubit lattice used in the experiment. Here, Q${}_1$ is the system, Q${}_2$-Q${}_5$ acts as the entangled environment, and Q${}_6$-Q${}_9$ serves as the perturbation of the environment. The interaction is realized with conditional gate $U_{\oslash}$. ({\bf B}) Schematic of the quantum circuit used to measure the main information theoretic quantities. Initially, Q${}_1$ is prepared in a superposition of $|0\rangle$ and $|1\rangle$ states by applying a Hadamard gate; subsequently, four $U_{\oslash}(2\theta)$ are applied to correlate the system and four environment qubits. Before measuring, the perturbation is added to the environment through the conditional gate  $U_{\oslash}(\theta+3\pi/2)$. ({\bf C}) Numerical and experimental results of measuring mutual information between the system and the environment. ({\bf D}) Experimental results of measuring Holevo bound $\chi$ (top) and quantum discord $\mathcal{D}$ (bottom). The markers (lines) are the experimental (numerical) results. Data points are measured over five independent runs, and error bars represent the standard deviations of these results.}
    \label{fig:fig3}
\end{figure*}

\noindent The essence of quantum Darwinism is understanding how the system-environment information exchange leads to the emergence of classicality through encoding copies of the classical information of $\cS$ in independent fragments of $\cE$. The only compatible form of the $\cal SE$ joint-state is the singly-branching form~\cite{touil2022branching}, i.e., the one that filters out the system's pointer states~\cite{blume2005simple,blume2006quantum}. To probe this structure of states, geometric quantum mechanics~\cite{Bengtsson_Zyczkowski_2006, fabio1, fabio2, fabio3} emerges as a powerful framework (supplementary text section 1B). Here, the quantum state space is formed by the complex projective Hilbert space $\mathcal{P}(\mathcal{H}) = \mathbb{C}P^{D-1}$ for a system with Hilbert space $\mathcal{H}$ of dimension $D$, whose geometric structure is characterized by an invariant measure, the Fubini-Study metric. In particular, a pure state — a point $Z_0 \in \mathcal{P}(\mathcal{H})$ — is represented by a Dirac measure $\mu_{\text{pure}} = \delta_{Z_0}$, while a mixed state corresponds to a complex combination of weighted Dirac measures, $\mu_{\text{mix}} = \sum_j \lambda_j \delta_{Z_j}$, with $\sum_j \lambda_j = 1$. Within this picture, any joint pure state of $\cS\cE$, undergoing decoherence, can be expressed as:
\begin{align}
\left|\psi_{\mathcal{SE}}\right\rangle &= \sum_{i, \alpha, \beta} \psi_{i \alpha \beta}\left|s_i\right\rangle\left|f_\alpha\right\rangle\left|\bar{f}_\beta\right\rangle \notag \\
&= \sum_{\alpha, \beta} \sqrt{X_{\alpha \beta}}\left|\chi_{\alpha \beta}\right\rangle\left|f_\alpha\right\rangle\left|\bar{f}_\beta\right\rangle\,,
\label{eq:join_state}
\end{align}
with $\ket{s_i},$ $\ket{f_\alpha}$ and $\ket{\bar{f}_\beta}$ being orthonormal states of $\cal S$, $\cal F$ and of the environment complement $\cal \bar F$, respectively; $X_{\alpha \beta}$ is the probability of $\cS\cE$ in the composite state $\left|\chi_{\alpha \beta}\right\rangle\left|f_\alpha\right\rangle\left|\bar{f}_\beta\right\rangle$. This representation visualizes the state as a measure on the projective Hilbert space $\mathcal{P}(\mathcal{H}_{\mathcal{S}})$, such that decoherence manifests as the geometric state of $\cS$, $\mu_{\cS}=\sum_{\alpha,\beta}X_{\alpha\beta}\delta_{\chi_{\alpha\beta}}$, begins to cluster around the pointer states.

For instance, let a system composed of a single-qubit $\cS$ interacting with an environment $\cE$ with $N$-qubits via the conditional gate~\cite{touil2022}, $U^{k}_{\oslash}=|0_{\mathcal{S}}\rangle\langle0_{\mathcal{S}}|\otimes U_{k}^{0}+|1_{\mathcal{S}}\rangle\langle1_{\mathcal{S}}|\otimes U_{k}^{1}$, where $|0_{\mathcal{S}}\rangle,~|1_{\mathcal{S}}\rangle$ are two orthogonal pointer states of $\mathcal{S}$, and the controlled unitary to the $k$-th environment qubit, $\mathcal{E}_k$, is $U_k^j= \exp[(-{\rm i}\theta_k^j/2)(\sigma_x\cos\phi_k^j+\sigma_y\sin\phi_k^j)]$. The randomly chosen parameters $\{\theta_k^j;\phi_k^j\} \in \{\left[(j-0.5)\pi, (j+0.5)\pi\right); \left[-\pi,\pi\right)\}$ quantify the imperfect encoding of the information about $\mathcal{S}$ in $\mathcal{E}$. Given $|0_{\mathcal{S}}\rangle=|0\rangle$ and $|1_{\mathcal{S}}\rangle=|1\rangle$, the form of the unitary leads $\mu_{\cS}$ to develop two clusters at antipodal points on the Bloch sphere for sufficiently large $N$ (Fig.~\ref{fig:fig1}D, top), as a result of decoherence, and indicates the einselection of stable pointer states in the macroscopic `classical world'. If instead the environment is subdivided into two disjoint fragments ${\cal F}_A$ and ${\cal F}_B$, the geometric quantum states, $\mu_{\cS}^{A}$ and $\mu_{\cS}^{B}$, equivalently cluster to two deterministic states at large $N$ (Fig.~\ref{fig:fig1}D, bottom), while large uncertainty arises at small $N$ due to the residual quantum coherence. The emergence of classicality now becomes clear: Two independent observers eavesdropping on separate environments agree on the measured information of the system.

Moving to experimental exploration, we utilize 12 qubits on our superconducting processor (supplementary text section 2E) to construct a slightly more complex scheme comprised of a system $\cS$ formed by \textit{two} central entangled qubits which are coupled to $N=10$ surrounding qubits simulating the photon environment $\cE$ (Fig.~\ref{fig:fig2}A); the `photons' only interact with the system and not with each other. Here, a similar randomized conditional gate $U^{k}_{\oslash}$ is defined with pointer states $|0_{\mathcal{S}}\rangle=|00\rangle$ and $|1_{\mathcal{S}}\rangle=|11\rangle$. In our experiments, $\cS$ is initialized to $\ket{\Psi_\mathcal{S}^0} = \frac{1}{\sqrt{2}}\left(\ket {00} + \ket {11} \right)$ by applying three Hadamard gates and a CZ gate (Fig.~\ref{fig:fig2}B). Subsequent application of conditional gates $\{U^{k}_{\oslash}\}$ correlate all environment qubits $\{\cE_k\}$ with $\mathcal{S}$, resulting in the branching state~\cite{Zurek2009NP}, 
\begin{equation}
\ket{ \Psi^{\oslash}_\mathcal{SE}} = \frac{1}{\sqrt{2}}\left(\ket {00}\bigotimes_{k=1}^{N} \ket {0_{\cE_{k}}} +   \ket{11} \bigotimes_{k=1}^{N} \ket {1_{\mathcal{E}_k}}\right)\,,
\label{bstate}
\end{equation}
where $\ket {j_{\cE_{k}}} = \cos(\theta_k^j/2)\ket{0^k}-{\rm i}\sin(\theta_k^j/2)e^{{\rm i}\phi_k^j}\ket{1^k}$ ($j=0,1$), recording the information about the pointer states on the $k$-th environment qubit  $\mathcal{E}_k$. See supplementary text section 1C for the analytical calculation of this model.

To experimentally measure the geometric state $\mu_\mc{S}$, we apply quantum state tomography pulses (see supplementary text sections 2B and 2C for details) on $\mc{S}$ before performing projective measurements on all qubits in the computational basis. Note that the circuit (Fig.~\ref{fig:fig2}B) is further compiled into combinations of  CZ gates and single-qubit rotations during execution (supplementary text section 2E). From the measurement outcomes, we reconstruct the ensemble realizations $\{X_{\alpha\beta}, \rho_{\cS}^{\alpha\beta}\}$ of $\mc{S}$ and record the corresponding basis $\left|f_\alpha\right\rangle\left|\bar{f}_{{\beta}}\right\rangle$ of the $N$-qubit environment, where $\rho_{\cS}^{\alpha\beta}$ is the density matrix of $|\chi_{\alpha\beta}\rangle$. The inset of Fig.~\ref{fig:fig2}C visualizes the measured  $\{X_{\alpha\beta}, \rho_{\cS}^{\alpha\beta}\}$ on the Bloch sphere of pointer states for different environment sizes $N=$ 2, 6, and 10. Notably, although the interactions between $\mc{S}$ and each composition of $\mc{E}$ are randomly sampled, a branching structure naturally arises through the decoherence induced by the growing size of the environment. When defining the integrated probability $P(\theta)$ of the states whose polar angles $\theta$ belong to the interval $[0, \theta]$, we obtain a direct estimation of the clustering of $\{X_{\alpha\beta}, \rho_{\cS}^{\alpha\beta}\}$ around the pointer states $|0_{\cS}\rangle$ and $|1_{\cS}\rangle$ (the two poles of the Bloch sphere, Fig.~\ref{fig:fig2}C). As the environment size $N$ grows, $P(\theta)$ becomes sharper at the two poles while leveling off at 0.5 for intermediate values of the polar angle $[\theta \in (0,\pi)]$. 
It is noteworthy that this observed self-organized
branching 
and losing of quantum coherence in our experiments is a pure effect of quantum unitary evolution, without any extra assumptions on measurements, which sheds light on settling the measurement problem, a fundamental postulate of quantum mechanics~\cite{basis1, Zurek2009NP}. 

Until now, all descriptions have been established in the quantum realm. A natural question then arises: How does the emerging branching structure of the globally pure wavefunction lead to classical reality? In quantum Darwinism, a key insight about the classical world is that the observer can only eavesdrop on {\it a fragment $\cF$ of the whole environment $\cE$} and deduce the information of ${\cS}$ from the recorded basis $\{\ket{f_\alpha}\}$ of $\cF$. To bridge the branching behavior and the information-theoretic signatures of quantum Darwinism, we focus on the system with an environment size $N=10$ and vary the fragment size $m$ of $\cF$. Figure~\ref{fig:fig2}D displays the numerical results of the mutual information $\MI$,  Holevo bound $\holevo$, and discord $\discord$, and exemplifies the experimentally measured geometric states $\{X_{\alpha}, \rho_{\cS}^{\alpha}\}$ of $\cS$ for $m=2,5,8$ (inset), which are chosen from the classical plateau $\MI\simeq1$. Here, $\rho_{\mathcal{S}}^{\alpha}=\sum_{\beta}\bra{\bar{f}_{\beta}}\bk{f_{\alpha}}{\Psi_\mathcal{SE}}\bk{\Psi_\mathcal{SE}}{f_{\alpha}}\ket{\bar{f}_{\beta}}$, is a mixed state due to a lack of information about $\bar{\cF}$, and $X_{\alpha}$ is the corresponding probability. Within the plateau regime, almost all correlations between $\cS$ and $\cF$ are classical, resulting in the observer having the ability to learn most of the shared information between $\cS$ and $\cF$ through measurements on $\cF$, apart from a measure-zero case where particular measurements reveal no information~\cite{Zwolak2013SR}. Therefore, $\holevo\simeq \MI$ and $\discord\simeq0$ at sufficiently large $m/N$ within the plateau. Correspondingly, $\{X_{\alpha}, \rho_{\cS}^{\alpha}\}$ gradually converge to two clouds around the pointer states $\{\ket{0_{\cS}}, \ket {1_{\mathcal{S}}}\}$ with a larger separation along the $z$-axis as $m$ grows. 

Another angle to show that the state ensemble $\{\rho_{\cS}^{\alpha}\}$ of $\cS$ is classically correlated with the bases $\{\ket{f_{\alpha}}\}$ ($f_{\alpha} = f_1^{\alpha} f_2^{\alpha}...f_m^{\alpha}, f_i^{\alpha}\in\{0, 1\}$) of $\cF$ that are eavesdropped by the observer is shown in Fig.~\ref{fig:fig2}E. Here, we report the measured signal of $\mathcal{A}(z)=\sum_{\{\alpha, \langle\sigma_z^{\alpha}\rangle=z\}} {X_{\alpha}}\sum_{i=1}^{m}{(1 - 2f_i^{\alpha})}$ along Bloch sphere's $z$-axis, where $\langle\sigma_z^{\alpha}\rangle=\Tr{(\rho_{\cS}^{\alpha}\sigma_z)}$. Remarkably, two branches also emerge in the distribution of $\mathcal{A}(z)$, establishing a one-to-one correspondence with the two pointer states of $\cS$: $\mathcal{A}<0\Rightarrow\ket{1_{\cS}}$, $\mathcal{A}>0\Rightarrow\ket{0_{\cS}}$. Thus, an observer can learn which pointer state the system $\cS$  `collapses' into by calculating $\mathcal{A}(z)$ from the eavesdropped classical bit string $f_\alpha$ imprinted on $\cF$. As $m$ increases, $\mathcal{A}(z)$ tends to congregate to $z=\pm1$ with higher signal amplitudes, allowing the observer more confidence to confirm the already known information. Therefore, extra data provided by larger fragments are redundant. As long as the observers eavesdrop on a suitably large $\cF$, they always agree on their conclusion if they are in the same branch. These observations illustrate how classical reality emerges from a structured quantum universe and builds up its connection with the classical plateau of mutual information. 

\vspace{.5cm}
\noindent\textbf{Decoherence and quantum Darwinism}
\vspace{.2cm}

\noindent To fully comprehend and test the quantum-to-classical transitions, we now measure the information-theoretic signatures of classicality in quantum Darwinism, i.e., the plateau of mutual information and the vanishing discord. 
For that, we now employ a slightly smaller circuit featuring nine qubits (Fig.~\ref{fig:fig3}A), with a single qubit in ${\cal S}$.  At the same time, we investigate a more generic quantum system where four environment qubits interact weakly via an extra set of four auxiliary ones, allowing for the verification of the emergence of classicality with the interplay of information scrambling in $\cE$~\cite{duruisseau2023pointer,PRXQuantum.2.010306,Touil_2024}. The couplings between $\cS$ and the four directly connected environment qubits are homogeneous and realized via a conditional gate $U_{\oslash}(\theta)=|0_{\cS}\rangle\langle0_{\cS}|\otimes\mathbb{I} +|1_{\cS}\rangle\langle1_{\cS}|\otimes \exp[-{\rm i}\theta\sigma_{y}/2]$, where the pointer states  $\ket{0_{\cS}}=\ket{0},~\ket{1_{\cS}}=\ket{1}$.

Similar to previous cases, the system $\mathcal{S}$ starts in a pure state $\ket{\Psi_\mathcal{S}^0}=\frac{1}{\sqrt{2}}(|0\rangle+|1\rangle)$ by applying a Hadamard gate, and the unitary $U_{\oslash}(2\theta)$ correlates each qubit in $\cE$ with $\cS$, resulting on a branching state, 
\begin{equation}
{\ket{ \Psi^{\oslash}_\mathcal{SE}} = \frac{1}{\sqrt{2}}\,  \left(\ket 0\bigotimes_{k=1}^{N} \ket{0_{\cE_{k}}} + \,  \ket{1}  \bigotimes_{k=1}^{N} \ket {1_{\mathcal{E}_k}}\right) \, .}
\end{equation}
In addition, we couple the four environment qubits through four auxiliary qubits with a similar unitary but different interaction `strength' $U_{\oslash}(\theta+3\pi/2)$. The full quantum circuit is shown in Fig.~\ref{fig:fig3}B. By varying $\theta \in [0,\pi]$, we experimentally measure the mutual information $\MI$ (Fig.~\ref{fig:fig3}C), the Holevo bound $\holevo$ (Fig.~\ref{fig:fig3}D), and the quantum discord $\discord$ (Fig.~\ref{fig:fig3}D, see supplementary text section 2D for measurement details) using full quantum state tomography for a single initialized state in the 9-qubit lattice (Fig.~\ref{fig:fig3}A). The novel experimental findings in Fig.~\ref{fig:fig3}C and D accomplish complete verification of the predictions of quantum Darwinism. 

In particular, we observe that when one expects the `classical' limit to set in (i.e., for $\theta \in [\pi/2-\epsilon, \pi/2+\epsilon]$ where the extra environment scrambling unitaries amount to identity operations, thus decoupled) the mutual information has a steep rise to the classical plateau region where $\MI\simeq H_{\cS}$ for environment fragment size $m=1$. Capturing more qubits does not change its value unless we include almost all of the environment in the fragment ($m=4$), in which case $\MI\to 2H_{\cS}$. The experimentally measured $\MI$ for  $m=4$ at $\theta=\pi/2$ is about 1.83. Additionally, in the same regime, the quantum discord is arbitrarily close to zero until we capture the whole environment, in which case we obtain a peak where $\discord\to H_{\cS}$ for $m=N=4$ and $\theta=\pi/2$, indicating that quantum correlations are a global property of the composite system. This directly confirms that the emergence of classicality is quite robust, even if small imperfections exist, e.g., weak environmental couplings. 

\begin{figure}[t!]
    \begin{center}
    \includegraphics[width=\linewidth]{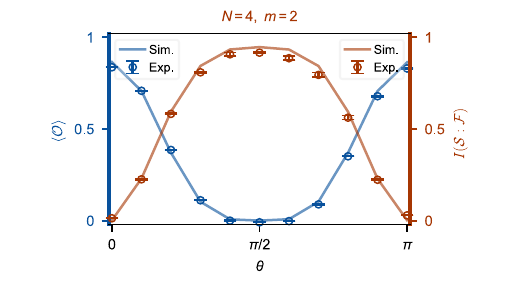}
    \end{center}
    \caption{\textbf{Witnessing quantum Darwinism with local observables.}
    The expectation value of $\mc{O}= \sigma_{x}\otimes\frac{1}{\sqrt{5}}(2\sigma_z+\sigma_{y})\otimes\mathbb{I}\otimes\mathbb{I}\ldots\mathbb{I}$ and mutual information $\MI$ as a function of $\theta\in [0, \pi]$. The circles (solid lines) represent experimental (numerical) results. The plateau at zero is established when $\theta$ is close to $\pi/2$. The mutual information between two environment qubits and the system averaged over all combinations for $m=2$, $N=4$. Measuring $\MI$ is time-consuming as full quantum state tomography is used. Data points are measured over five runs, and error bars represent the standard deviations of these five results.}
    \label{fig:fig4}
\end{figure}
 
\vspace{.5cm}
\noindent\textbf{Local observables}
\vspace{.2cm}

\noindent Despite the success of probing branching structures and information-theoretic measures to interpret the origin of classical reality, obtaining them for large systems is experimentally and numerically prohibitive due to unscalable quantum tomography and matrix diagonalization. However, the insight of a highly structured quantum wavefunction provided by geometric quantum mechanics inspires us to propose a new quantifier to witness quantum Darwinism. Let $\mathcal{O}$ given by 
\begin{equation}
\mathcal{O}=A\otimes B \otimes \mathbb{I}\otimes \cdots \otimes \mathbb{I}\,,
\end{equation}
where $A$ acts on the system $\cS$ and $B$ acts on a fragment $\cF$ of the environment. We note that $\mathcal{O}$ is {\it local} (i.e., a few-body operator) with a minimal structure, which we will argue to be able to build a one-to-one correspondence between the vanishing of its expectation value and the emergence of a branching structure. For example, if the system exhibits the set of pointer states $\{|n\rangle\}$, one can choose $A$ such that it rotates a pointer state to an orthogonal one (e.g., $A|n\rangle \rightarrow |n+1\rangle$). Additionally, we must choose an operator $B$ such that $\langle \mathcal{O}\rangle \neq 0$ when the composite state $\left|\psi_{\mathcal{SE}}\right\rangle$ is not in a branching structure. This minimal form of $\mathcal{O}$ is necessary because observables acting solely on the system may not capture the essential dynamics associated with the branching structures. As a result of this form, $\langle\mathcal{O}\rangle$ becomes arbitrarily close to zero as long as the system approaches a branching form. This behavior provides a direct and inexpensive method to detect the emergence of a branching structure and the plateau characteristic of quantum Darwinism. In supplementary text section 1D, we provide analytical proof of the behavior of such local observables and further argue for the necessity of the existence of $B$. 

Turning to the experiments, for the circuit in Fig.~\ref{fig:fig3}B, instead of tomography, we directly measure the expectation of the observable $\mathcal{O} = \sigma_{x}\otimes (1/\sqrt{5})(2\sigma_z+\sigma_{y})\otimes\mathbb{I}\otimes\mathbb{I}\ldots\mathbb{I}$, see Fig.~\ref{fig:fig4}. Here, $\langle {\cal O}\rangle$ is measured for $\theta$ between 0 and $\pi$ for an environmental size $N=4$; in comparison, the mutual information $\MI$ is measured for a fragment size $m=2$. We observe a one-to-one correspondence with the zero-plateau of $\langle\mathcal{O}\rangle$ and the emergence of the branching structure as captured by the plateau of $\MI$, demonstrating a convenient and inexpensive way to detect the emergent classical reality.

\vspace{.5cm}
\noindent\textbf{Discussion and outlook}
\vspace{.2cm}

\noindent Understanding the principles of quantum mechanics remains a significant challenge in physics due to the inherently non-intuitive nature of quantum phenomena, including that of the measurement problem. 
Here, we presented a robust experimental verification of the predictions of quantum Darwinism~\cite{darwin1,darwin2,darwin3,blume2005simple,darwin5,darwin6,darwin7,darwin8}, a physical framework that has the merit of addressing such foundational divide between quantum and classical worlds. At its core is the formation of a branching structure of the global state promoted by the decoherence of a system of interest under the action of its witnessing environment. The tools of geometric quantum mechanics allow, for the first time, the direct observation of branching and the resulting clustering around the system's pointer states, which supports the emergence of classical reality. 

Addressing previous limitations of experimental efforts, our investigation further shows that carefully chosen yet simple quantifiers, such as certain local observables, can be used for probing quantum-to-classical transitions. With quantum Darwinism now experimentally validated and established as a mature field of research, we can explore its rich potential in addressing further issues of the quantum measurement problem, such as the dynamics of the collapse of the wave function. One potential avenue is studying the interplay between the emergence of classicality and thermalization in open quantum systems. Our findings thus pave the way for connecting two of the most successful fields in physics: thermodynamics and quantum theory.\\

\vspace{.5cm}
\noindent\textbf{ACKNOWLEDGMENTS}
\vspace{.2cm}

\noindent A.T. would like to thank Wojciech H. Zurek for insightful discussions. The device was fabricated at the Micro-Nano Fabrication Center of Zhejiang University. 
R.M. acknowledges that part of the calculations used resources from the Research Computing Data Core at the University of Houston.
{\bf Funding:} The experimentation team acknowledges the support from the National Natural Science Foundation of China (Grant Nos. 12274368, 12174342, 12274367, 12322414, 12404570, 12404574, U20A2076), the Zhejiang Provincial Natural Science Foundation of China (Grant Nos. LR24A040002 and LDQ23A040001).
A.T. acknowledges support from the U.S. DOE under the LDRD program at Los Alamos.
R.M.~acknowledges support from the T$_{\rm c}$SUH Welch Professorship Award.
{\bf Author contributions:} Z.Z. carried out the experiments and analyzed the experimental data under the supervision of Q.G.; 
Z.Z., K.S., and Z.B. performed numerical simulation under the supervision of R.M. and Q.G.; 
R.M., K.S., A.T. and Q.G. conducted the theoretical analysis for the experiments. 
H.L. fabricated the device under the supervision of H.W.;
K.S., A.T., Q.G, R.M. and Z.Z. co-wrote the manuscript and
Z.Z., Z.B., Z.S., P.Z., H.L., Z.W., C.S., Q.G., and H.W. contributed to the experimental setup. 
All authors contributed to the discussions of the results.
{\bf Competing interests:} The authors declare no competing interests. 
{\bf Data and materials availability:} The data presented in the figures and that support the other findings of this study will be publicly available upon its publication. All the relevant source codes are available from the corresponding authors upon reasonable request.

\vspace{.5cm}
\noindent\textbf{SUPPLEMENTARY MATERIALS}
\vspace{.2cm}

\noindent Supplementary Text\\
Figs. S1 to S2\\
Table S1 \\
References~(S1-S16)\\

\bibliography{main}

\end{document}


\title{Supplementary Materials:\\
Observation of Quantum Darwinism and the Origin of Classicality with Superconducting Circuits
}
\maketitle
\onecolumngrid

\tableofcontents
\beginsupplement

\section{Theoretical analysis}
\label{supp}
\subsection{Decoherence theory and pointer states}
Decoherence theory addresses how quantum systems interacting with their environment effectively lose their quantum coherence, leading to the destruction of quantum superpositions. When a quantum system $\mc{S}$ interacts with its environment $\mc{E}$, the total system can be described by a Hamiltonian,
\begin{equation}
H = H_{\rm sys.} + H_{\rm env.} + H_{\rm int.},
\end{equation}
where $H_{\rm sys.}$ and $H_{\rm env.}$ are the self-Hamiltonians of the system and environment, respectively, and $H_{\rm int.}$ represents their interaction. The unitary evolution under this interacting Hamiltonian suffices to build up quantum entanglement over $\cS\cE$, leading to non-separable correlations between $\cS$ and $\cE$. In this scenario, only a set of preferred states, called pointer states $\{|n_{\mc{S}}\rangle\}_n$, are einselected, remaining unchanged under the system-environment interactions~\cite{basis1,basis2}, while other states are decohered into mixtures of them. Specifically, if an observable to be measured commutes with the total Hamiltonian \( H \) such that 
\begin{equation}
[H, O_{\mc{S}} \otimes \mathbb{I}_{\mc{E}}] = 0,
\end{equation}
this observable is conserved under the dynamics governed by \( H \) and will not be perturbed by the system-environment interaction. In the representation of pointer states, the off-diagonal elements (quantum coherences) of the system's reduced density matrix decay rapidly in the limit of good decoherence, effectively suppressing quantum interference effects. In particular, after decoherence, we obtain,
\begin{align}
\rho_{\cS} &= \mathrm{Tr}_{\cE} (\rho_{\mc{SE}}) = \sum_{n} p_{n} \ket{n_{\mc{S}}} \bra{n_{\mc{S}}},\\
\langle O_{\cS}\rangle &= \mathrm{Tr} (\rho_{\mc{S}}O_{\mc{S}}) = \sum_{n} p_{n} \langle n_{\mc S}| O_{\mc{S}} \ket{n_{\mc{S}}}\ .
\end{align}
Another method to determine the pointer states, known as the `predictability sieve'~\cite{Zurek1993a,Zurek1993}, involves selecting states that minimize the entropy production due to decoherence, effectively picking states that retain maximal predictability over time. Mathematically, pointer states $\{|n_{\mc{S}}\rangle\}_n$ are the states whose von Neumann entropy $H(\rho_{\cS}) = -\mathrm{Tr}[\rho_{\cS} \log_2( \rho_{\cS})]$ is constant over time. When the self-Hamiltonian of the system does not commute with the interaction Hamiltonian the notion of pointer states can be only approximate and in such cases we either look for observables that almost commute with the Hamiltonian $[H, O_{\mc{S}} \otimes \mathbb{I}_{\mc{E}}] \approx 0$, or the states that have minimal changes in their von Neumann entropy as a function of time.

\subsection{Geometric quantum mechanics}
As detailed in Ref.~\cite{fabio3, Touil2022branching}, Geometric Quantum Mechanics provides a formal framework for representing quantum states in the complex projective Hilbert space, defined as $\mathcal{P}(\mathcal{H}) = \mathbb{C}P^{D-1}$, instead of the conventional Hilbert space representation for a system with dimension $D$. This geometric framework has proven instrumental in developing intuition about quantum many-body dynamics. In contrast to the description of density matrices, geometric quantum states are capable of expressing the specific realization of a mixed-state ensemble. More specifically, based on a chosen set of orthonormal basis $\{\left|e_\alpha\right\rangle\}_\alpha$ of the Hilbert space, a pure state can be expressed in terms of $D$ complex coordinates as $|Z\rangle = \sum_{\alpha} Z^\alpha \left|e_\alpha\right\rangle$, with the equivalence relation $Z \sim \lambda Z$ for all $\lambda \in \mathbb{C} \setminus \{0\}$ ensuring $Z \in \mathcal{P}(\mathcal{H})$. The state space $\mathcal{P}(\mathcal{H})$ is equipped with differential-geometric tools such as the Fubini-Study volume element. This preferred invariant measure facilitates the use of measure theory for defining ensembles and describing mixed states. Parameterizing pure states using probability-phase coordinates $Z^\alpha = \sqrt{p_\alpha} e^{i \nu_\alpha}$, the Fubini-Study volume element is expressed as $dV_{FS} \sim \prod_{\alpha=0}^{D-1} dp_\alpha \, d\nu_\alpha$, up to an overall normalization factor. The Fubini-Study volume element naturally defines a geometric quantum state as a probability measure $\mu$ on $\mathcal{P}(\mathcal{H})$, representing an ensemble of pure states. As we detail in the main text, a pure state $Z_0 \in \mathcal{P}(\mathcal{H})$ corresponds to a Dirac measure $\mu_{\text{pure}} = \delta_{Z_0}$, while a finite ensemble is represented as a convex combination of weighted Dirac measures: $\mu_{\text{ensemble}} = \sum_j \lambda_j \delta_{Z_j}$, with $\sum_j \lambda_j = 1$. The connection to density matrices then arises naturally: The elements $\rho_{\alpha \beta}$ of the density matrix are the expectation values of $Z^\alpha \bar{Z}^\beta$ with respect to $\mu$, given by $\rho_{\alpha \beta} = \mathbb{E}_\mu[Z^\alpha \bar{Z}^\beta] = \int_{\mathcal{P}(\mathcal{H})} d\mu(Z) Z^\alpha \bar{Z}^\beta$. In Ref.~\cite{fabio1, Touil2022branching}, this approach was used to derive the geometric quantum state of an open quantum system coupled with an environment. We use this framework to experimentally probe the geometric states of the system of interest $\cS$, and verify how the emergence of classicality reflects these states' structure. More precisely, the joint pure state of the system and environment $\left|\psi_{\mathcal{SE}}\right\rangle$ can be represented as
\begin{equation}
\begin{aligned}
\left|\psi_{\mathcal{S E}}\right\rangle \equiv\left|\psi_{\mathcal{S F} \bar{\mathcal{F}}}\right\rangle & =\sum_{i, \alpha, \beta} \psi_{i \alpha \beta}\left|s_i\right\rangle\left|f_\alpha\right\rangle\left|\bar{f}_\beta\right\rangle, \\
& =\sum_{\alpha, \beta} \sqrt{X_{\alpha \beta}}\left|\chi_{\alpha \beta}\right\rangle\left|f_\alpha\right\rangle\left|\bar{f}_\beta\right\rangle,
\end{aligned}
\end{equation}
which can be interpreted as a measure on the projective Hilbert space $\mathcal{P}(\mathcal{H}_{\mathcal{S}})$. Decoherence then leads to clustering of the states $\ket{\chi_{\alpha\beta}}$ around pointer states. For instance, in a system comprised of a single qubit with pointer states $\left|0\right\rangle$ and $\left|1\right\rangle$, this clustering results in geometric quantum states forming two distinct clusters at antipodal points on the Bloch sphere, which is corroborated by the experimental results presented in the main manuscript.

\subsection{Analytical calculation of the mutual information}
We provide an in-depth calculation of the mutual information, $\MI=H_{\cS}+H_{\cF}-H_{\cS\cF}$, for the system-environment interaction model used in Fig.~2 of the main text, where the system $\cS$ interacts with an $N$-qubit environment $\cE$ via conditional gates, $U^{k}_{\oslash}=|0_{\mathcal{S}}\rangle\langle0_{\mathcal{S}}|\otimes U_{k}^{0}+|1_{\mathcal{S}}\rangle\langle1_{\mathcal{S}}|\otimes U_{k}^{1}$, to show that it has a closed expression as studied in Ref.~\cite{touil2022}. We will be interested in the correlations between the fragment $\cF$ with a size $m$ and the system $\cal S$. The system starts in the state superposition $|\Psi_{\mc{S}}^0\rangle= \sqrt{p}|0_{\cS}\rangle+ \sqrt{q}|1_{\cS}\rangle$, where the normalization coefficients $p$ and $q$ ($p=q=1/2$ in the main text) satisfy the relation $p+q=1$, while the environment starts in the ground state $|00...0\rangle$. In the main text, we show that without considering auxiliary qubits, the resulting branching state is
\begin{align}
    \ket{ \Psi^{\oslash}_\mathcal{SE}} = \sqrt{p}\ket {0_{\cS}}\bigotimes_{k=1}^{N} \ket {0_{\cE_{k}}} + \sqrt{q}\ket{1_{\cS}} \bigotimes_{k=1}^{N} \ket { 1_{\mathcal{E}_k}}, \label{eq:branch}
\end{align}
where $\ket {j_{\cE_{k}}} = \cos(\theta_k^j/2)\ket{0^k}-{\rm i}\sin(\theta_k^j/2)e^{{\rm i}\phi_k^j}\ket{1^k}$ ($j=0,1$). $\{\ket{j_{\cE_k}}\}$ composes a set of non-orthogonal basis. Hence the overlap between $|0_{\cE_k}\rangle$ and $|1_{\cE_k}\rangle$  is
\begin{align}
    s_{k} = \langle 1_{\cE_k}|0_{\cE_k}\rangle = \cos(\theta_k^{1}/2)\cos(\theta_k^{0}/2)+e^{i(\phi_k^0-\phi_k^{1})}\sin(\theta_k^{1}/2)\sin(\theta_k^{0}/2). \label{eq:overlap}
\end{align}
We can define
\begin{equation}
\begin{split}
    &|\mc{F}_{0}\rangle = \bigotimes_{k=1}^{m}|0_{\mc{E}_{k}}\rangle \ \text{and}\ |\bar{\mc{F}_{0}}\rangle = \bigotimes_{k=m+1}^{N}|0_{\mc{E}_{k}}\rangle,\\
     &|\mc{F}_{1}\rangle = \bigotimes_{k=1}^{m}|1_{\mc{E}_{k}}\rangle \ \text{and} \ |\bar{\mc{F}}_{1}\rangle = \bigotimes_{k=m+1}^{N}|1_{\mc{E}_{k}}\rangle.
\end{split}
\end{equation}
Now, we can express the wave function as
\begin{equation}
\begin{aligned}
    |\Psi_{\mc{SE}}^{\oslash}\rangle &= \sqrt{p}|0_\cS\rangle|\mc{F}_{0}\rangle|\bar{\mc{F}}_{0}\rangle +\sqrt{q}|1_\cS\rangle|\mc{F}_{1}\rangle|\bar{\mc{F}}_{1}\rangle\\
    &= \sqrt{p}|0_\cS\rangle|\mc{E}_{0}\rangle+\sqrt{q}|1_\cS\rangle|\mc{E}_{1}\rangle,
\end{aligned}
\end{equation}
and the overlaps as
\begin{align}
    &\langle\mc{E}_{1}|\mc{E}_{0}\rangle = \prod_{k=1}^{N} s_{k},\\
    &\langle\mc{F}_{1}|\mc{F}_{0}\rangle = \prod_{k=1}^{m} s_{k},\label{eq:overlap_FF}\\
    \text{and}\ \ &\langle\bar{\mc{F}_{1}}| \bar{\mc{F}}_{0}\rangle = \prod_{k=m+1}^{N} s_{k}.
\end{align}
We then see that the density matrix $\rho_{\mc{S}}$ is
\begin{equation}    
\begin{aligned}
    \rho_{\mc{S}}=&\Tr{_\mc{E}}(|\Psi_{\mc{SE}}^{\oslash}\rangle\langle\Psi_{\mc{SE}}^{\oslash}|)\\
    =&\sum_{i} \langle i| \left[p|0_\cS\rangle\langle 0_\cS|\bigotimes|\mc{E}_{0}\rangle \langle \mc{E}_{0}|
    + q|1_\cS\rangle\langle 1_\cS|\bigotimes|\mc{E}_{1}\rangle \langle \mc{E}_{1}|
    + \sqrt{pq}|0_\cS\rangle\langle 1_\cS|\bigotimes|\mc{E}_{0}\rangle \langle \mc{E}_{1}|
    + \sqrt{pq}|1_\cS\rangle\langle 0_\cS|\bigotimes|\mc{E}_{1}\rangle \langle \mc{E}_{0}|\right]|i\rangle
    \\
    =&\sum_{i} \big[p|0_\cS\rangle\langle0_\cS|\langle i|\mc{E}_{0}\rangle \langle\mc{E}_{0}|i\rangle+q|1_\cS\rangle\langle1_\cS|\langle i|\mc{E}_{1}\rangle \langle\mc{E}_{1}|i\rangle+\sqrt{pq}|0_\cS\rangle\langle1_\cS|\langle i|\mc{E}_{0}\rangle \langle\mc{E}_{1}|i\rangle+ \sqrt{pq}|1_\cS\rangle\langle0_\cS|\langle i|\mc{E}_{1}\rangle \langle\mc{E}_{0}|i\rangle\big]\\
    =&p|0_\cS\rangle\langle0_\cS|\left(\sum_{i}\langle \mc{E}_{0} | i \rangle \langle i | \mc{E}_{0} \rangle\right) 
    + q|1_\cS\rangle\langle1_\cS|\left(\sum_{i}\langle \mc{E}_{1} | i \rangle \langle i | \mc{E}_{1} \rangle\right) \\
    &+ \sqrt{pq}|0_\cS\rangle\langle1_\cS|\left(\sum_{i}\langle \mc{E}_{1} | i \rangle \langle i | \mc{E}_{0} \rangle\right) 
    +\sqrt{pq}|1_\cS\rangle\langle0_\cS|\left(\sum_{i}\langle \mc{E}_{0} | i \rangle \langle i | \mc{E}_{1} \rangle\right),
\end{aligned}
\end{equation}
where $\ket{i}\in\{\bigotimes_{k=1}^N\ket{j^k},j=0,1\}$. Since the basis set $\{\ket{i}\}$ is orthonormal satisfying $\sum_{i}|i\rangle\langle i| = \mathbb{I}$, we get
\begin{equation}
\begin{aligned}
\rho_{\mc{S}}& = p|0_\cS\rangle\langle0_\cS|+ q|1_\cS\rangle\langle1_\cS|+\sqrt{pq}|0_\cS\rangle\langle1_\cS|\langle \mc{E}_{1}|\mc{E}_{0}\rangle  + \sqrt{pq}|1_\cS\rangle\langle0_\cS|\langle \mc{E}_{0}|\mc{E}_{1}\rangle\\
& = \begin{bmatrix}
          p & \sqrt{pq}\langle\mc{E}_{1}|\mc{E}_{0}\rangle\\
          \sqrt{pq}\langle\mc{E}_{0}|\mc{E}_{1}\rangle&  q
      \end{bmatrix}\ \\
& = \begin{bmatrix}
          p & \sqrt{pq}\prod\limits_{k=1}^{N}s_k\\
          \sqrt{pq}\prod\limits_{k=1}^{N}s_k^*&  q
      \end{bmatrix}.\label{eq:rhoS}
\end{aligned}
\end{equation}
Now, the density matrix $\rho_{\mc{SF}}$ of the composite system $\mc{SF}$ directly follows Eq.~\eqref{eq:rhoS} as
\begin{align}
    \rho_{\mc{SF}}=&\begin{bmatrix}
                  p & \sqrt{pq}\prod\limits_{k=m+1}^{N}s_k\\
                  \sqrt{pq}\prod\limits_{k=m+1}^{N}s_k^*&  q
              \end{bmatrix}.\label{eq:rhoSF}
\end{align}
Similarly, we get 
\begin{align}
          \rho_{\cS\bar{\cF}}=\begin{bmatrix}
          p & \sqrt{pq}\prod\limits_{k=1}^{m}s_k\\
          \sqrt{pq}\prod\limits_{k=1}^{m}s_k^*&  q
      \end{bmatrix}.\label{eq:rhoF}
\end{align}
Thus, we can calculate $H_{\cS}$, $H_{\cS\cF}$, and $H_{\cS\bar{\cF}}$ efficiently because $\rho_{\cS}$, $\rho_{\cS\cF}$, and $\rho_{\cS\bar{\cF}}$ have been reduced to rank-two density matrices. In Eq.~\eqref{eq:rhoSF}, the basis set of $\rho_{\cS\cF}$ is $\{\ket{0_\mathcal{S}} \ket{\mathcal{F}_0}, \ket{1_\mathcal{S}} \ket{\mathcal{F}_1}\}$, which is an orthogonal one due to the existence of pointer basis $\{\ket{0_\mathcal{S}}, \ket{1_\mathcal{S}}\}$. Note that the similar condition also holds for $\rho_{\cS}$ and $\rho_{\cS\bar{\cF}}$ but not for $\rho_{\cF}$, since $\ket{{\cF}_0}$ and $\ket{{\cF}_1}$ are not necessarily orthogonal [see Eq.~{\eqref{eq:overlap_FF}}], which makes the calculation of $H_{\cF}$ not easy. Fortunately, 
$\ket{\Psi_{\mc{SE}}^{\oslash}}$ is a pure state by definition, thus, by the symmetry of the von Neumann entropy, we have $H_{\mc{F}}=H_{\mc{S}\bar{\mc{F}}}\cdot$. 

Hence, $\MI$ can be computed exactly~\cite{touil2022} regardless of the sizes of $\cE$ and $\cF$:
\begin{equation}
{I}(\mathcal{S}:\mathcal{F})={h}(\lambda^{+}_{1,N,p})+{h}(\lambda^{+}_{1,m,p})-{h}(\lambda^{+}_{m+1,N,p}),
\label{mutinfo}
\end{equation}
where ${h}(x)=-x \log_2(x)-(1-x) \log_2(1-x)$, and $\lambda^{\pm}_{a,b,p}$ are the eigenvalues of the density matrices $\rho_{\cS}, \rho_{\cS\bar{\cF}}$, and $\rho_{\cS\cF}$, given by
\begin{equation}
\lambda^{\pm}_{a,b,p}=\frac{1}{2}\left(1 \pm \sqrt{\left(2p-1\right)^{2}+4p(1-p)\Pi^{b}_{k=a}|s_{k}|^{2}}\right)\,.
\label{root}
\end{equation}

To get more insight from the results above, we investigate $\rho_{\cS}$, $\rho_{\cS\bar{\cF}}$, $\rho_{\cS\cF}$ for a special case: $\theta_k^j=\theta^j,\phi_k^0-\phi_k^1=0$, in which they can be reduced to
\begin{align}
    \rho_{\cS} &=\begin{bmatrix}
            p & \sqrt{pq}\cos^{N}(\frac{\Delta\theta}{2})\\
            \sqrt{pq}\cos^{N}(\frac{\Delta\theta}{2})& q
    \end{bmatrix},\label{eq:rho_s_reduce}\\
    \rho_{\cS\bar{\cF}} &=\begin{bmatrix}
        p & \sqrt{pq}\cos^{m}(\frac{\Delta\theta}{2})\\
        \sqrt{pq}\cos^{m}(\frac{\Delta\theta}{2})& q
    \end{bmatrix},\label{eq:rho_f_reduce}\\
    \rho_{\cS\cF} &=  \begin{bmatrix}
        p & \sqrt{pq}\cos^{N-m}(\frac{\Delta\theta}{2})\\
        \sqrt{pq}\cos^{N-m}(\frac{\Delta\theta}{2})& q
    \end{bmatrix},\label{eq:rho_sf_reduce}
\end{align}
where $\Delta\theta=\theta^0-\theta^1$. We can use Eqs.~\eqref{eq:rho_s_reduce} - \eqref{eq:rho_sf_reduce} to interpret the behavior of $\MI$ displayed in Figs.~1B, 1C and 2D of the main text. Assuming $\Delta\theta\neq0$ and $p=q=1/2$, in the large $N$ limit, $H_{\cS}\rightarrow1$ since $\cos^N(\Delta\theta/2)\rightarrow0$, indicating that $\cS$ is fully decohered. In this case, $\MI\approx1+H_{\cF}-H_{\cS\cF}$. When $m\ll N$, the off-diagonal terms still exist for  $\rho_{\cS\bar{\cF}}$ but vanish for $\rho_{\cS\cF}$, so $\MI\approx1+H_{\cF}-1=H_{\cF}$. In this regime, the observer can only collect little information about $\cS$ by measuring $\cF$. As $m$ increases, $H_{\cF}\to1$ rapidly, caused by the stronger decoherence of $\rho_{\cS\bar{\cF}}$, which indicates the onset of the plateau of $\MI$ and the emergent classical reality. Furthermore, as $m$ grows close to $N$, $H_{\cS\cF}$ begins to decrease, leading to the mutual information $\MI=1+1-H_{\cS\cF}\to2$. In such a scenario, all the quantum information is stored in $\rho_{\cS\cF}$.

\subsection{Local observables as quantifiers for quantum Darwinism}
Now shifting focus to local observables, we show the full mathematical proof for how a chosen set of observables serves as a quantifier for classicality. First, we start with the limiting case of the singly-branching structures.

\paragraph*{Singly-branching states of arbitrary dimension:}

In Ref.~\cite{Touil2022branching}, the following theorem was derived:

\textit{Given a pure state $\ket{\psi_{\mathcal{S}\mathcal{F}\bar{\mathcal{F}}}}$ such that $\discord\leq \epsilon_\mathcal{D}$ and $|\MI-H_{\mathcal{S}}|\leq \epsilon_I$, then for all $\epsilon_{\mathcal{D}},\epsilon_I > 0$ there exists $\eta(\epsilon_{\mathcal{D}},\epsilon_I)\geq 0$ with $\eta \in \mathcal{O}(\epsilon_{\mathcal{D}},\epsilon_I)$ and a branching state $\ket{\mathrm{GHZ}}= \sum_{n=1}^{D_\mathcal{S}} \sqrt{y_n}\ket{n} \ket{\mathcal{F}_n}\ket{\bar{\mathcal{F}}_n}$ such that}
\begin{equation}
    \left\vert \langle \psi_{\mathcal{S}\mathcal{F}\bar{\mathcal{F}}} \vert \mathrm{GHZ}\rangle  \right\vert^2 \geq 1-\eta(\epsilon_{\mathcal{D}},\epsilon_I)
    .
\end{equation}
Essentially this postulates that the states that can support classicality are arbitrarily close to a GHZ state. 
Therefore, let's consider the following structure of states:
\begin{equation}
|\psi_{\mathcal{SE}}\rangle= \sum_{n=1}^{D_\mathcal{S}} \sqrt{p_n}\ket{n} \ket{\mathcal{F}_n}\ket{\bar{\mathcal{F}}_n},
\end{equation}
such that the states {$|n\rangle$} form the pointer basis, and the states $\ket{\bar{\mathcal{F}}}$ form an orthonormal basis, while $\ket{\mathcal{F}}$ are not necessarily orthonormal. The physical assumption here is that the fragment $\mathcal{F}$ is small while the rest of the environment $\bar{\mathcal{F}}$ is large enough to ensure orthogonality. Consider now the local observable:
\begin{equation}
\mathcal{O}=A\otimes B \otimes \mathbb{I}\otimes ... \otimes \mathbb{I}.
\end{equation}
Taking the expectation value with consideration of the branching states leads to
\begin{equation}
\langle \psi_{\mathcal{SE}} | \mathcal{O} | \psi_{\mathcal{SE}} \rangle = \sum^{D_{\mathcal{S}}}_{n=1} p_n \langle n | A | n \rangle \langle \mathcal{F}_n | B | \mathcal{F}_n \rangle.
\end{equation}
This is a valid approximation since the rest of the environment is large. The above expectation value tends to zero if $\langle n | A | n \rangle=0$. In fact, if we choose the observable $A$ such that it rotates the pointer basis $A|n\rangle \rightarrow |n+1\rangle$; e.g. the operation that does this is $\sigma_x$ if the pointer states are the eigenstates of $\sigma_z$: $|0\rangle$ and $|1\rangle$. With this choice of observable, we guarantee that whenever the wave function $|\psi_{\mathcal{SE}}\rangle$ tends to a branching form, the expectation of $\mathcal{O}$ is arbitrarily close to 0. One immediate question one might ask is the following: \textit{Why is this the minimal structure of local observables?} If we only consider the observable
\begin{equation}
\mathcal{O}=A\otimes \mathbb{I}\otimes ... \otimes \mathbb{I},
\end{equation}
then
\begin{equation}
\langle \psi_{\mathcal{SE}} | \mathcal{O} | \psi_{\mathcal{SE}} \rangle \approx \sum^{D_{\mathcal{S}}}_{n=1} p_n \langle n | A | n \rangle.
\end{equation}
For equal probabilities, we get
\begin{equation}
\langle \psi_{\mathcal{SE}} | \mathcal{O} | \psi_{\mathcal{SE}} \rangle = (1/D_{\mathcal{S}}) \sum^{D_{\mathcal{S}}}_{n=1} \langle n | A | n \rangle = (1/D_{\mathcal{S}})  \Tr(A),
\end{equation}
hence, any traceless observable would lead to zero expectation value when the plateau forms. Therefore, the minimal structure needed is the following:
\begin{equation}
\mathcal{O}=A\otimes B \otimes \mathbb{I}\otimes ... \otimes \mathbb{I}.
\end{equation}
Where the operator $B$ has support over a single qubit's Hilbert space. Now, we explicitly show the general form of the expectation value of the above local observables for arbitrary states $|\psi\rangle$. In particular, on the Hilbert space of system, fragment, and complementary fragment, we have,
\begin{equation}
\begin{split}
|\psi\rangle &= \sum_{i,j,k} \sqrt{q_{i,j,k}} |i\rangle |F_{j}\rangle |\bar{\mathcal{F}}_{k}\rangle\\
&=  \sum_{n,j} \sqrt{q_{n,j,n}} |n\rangle |F_{j}\rangle |\bar{\mathcal{F}}_{n}\rangle + \sum_{i\neq k,j} \sqrt{q_{i,j,k}} |i\rangle |F_{j}\rangle |\bar{\mathcal{F}}_{k}\rangle    \\
&= \sqrt{1-r} \left(\sum_{n} \sqrt{p_n} |n\rangle |\mathcal{F}_{n}\rangle |\bar{\mathcal{F}}_{n}\rangle\right) + \sqrt{r} |\phi\rangle\\
&= \sqrt{1-r} |\psi_{\mathcal{SE}}\rangle + \sqrt{r} |\phi\rangle\\
\end{split}
\end{equation}
and then by definition $\langle\phi| \psi_{\mathcal{SE}}\rangle=0$. From this general decomposition (such that $r \in [0,1]$) we see,
\begin{equation}
\langle \mathcal{O} \rangle = (1-r) \langle \psi_{\mathcal{SE}} | \mathcal{O} | \psi_{\mathcal{SE}} \rangle + r \langle \phi | \mathcal{O} | \phi \rangle+ 2 \sqrt{r(1-r)} {\rm Re}(\langle \psi_{\mathcal{SE}} | \mathcal{O} | \phi \rangle)\\
\end{equation}
Similar to before, if we choose the observable $A$ such that $A|n\rangle \rightarrow |n+1\rangle$ we get,
\begin{equation}
\langle \mathcal{O} \rangle= r \langle \phi | \mathcal{O} | \phi \rangle\\ + 2 \sqrt{r(1-r)} {\rm Re}(\langle \psi_{\mathcal{SE}} | \mathcal{O} | \phi \rangle)
\end{equation}
which we can choose to be nonzero by picking the appropriate operator $B$. Hence, for the general case ($r \neq 0$), we have a nonzero expectation value. Only when we are close to a branching structure can we guarantee that $\langle \mathcal{O} \rangle$ is equal to zero. From the theorem derived in~\cite{Touil2022branching}, the expectation value we are proposing not only captures the plateau in the mutual information but also captures `epsilon' discord. Thus, if we observe that $\langle \mathcal{O} \rangle$ is at a zero plateau, we guarantee that the mutual information is $\delta$ within a plateau and discord is less or equal than $\epsilon$, such that $\delta$ and $\epsilon$ are arbitrarily small as guaranteed by the theorem derived in Ref.~\cite{Touil2022branching}.


\section{Experimental information\label{sec:exp}}
\subsection{Device information}
\begin{figure}[h]
    \begin{center}
    \includegraphics[width=1\textwidth]{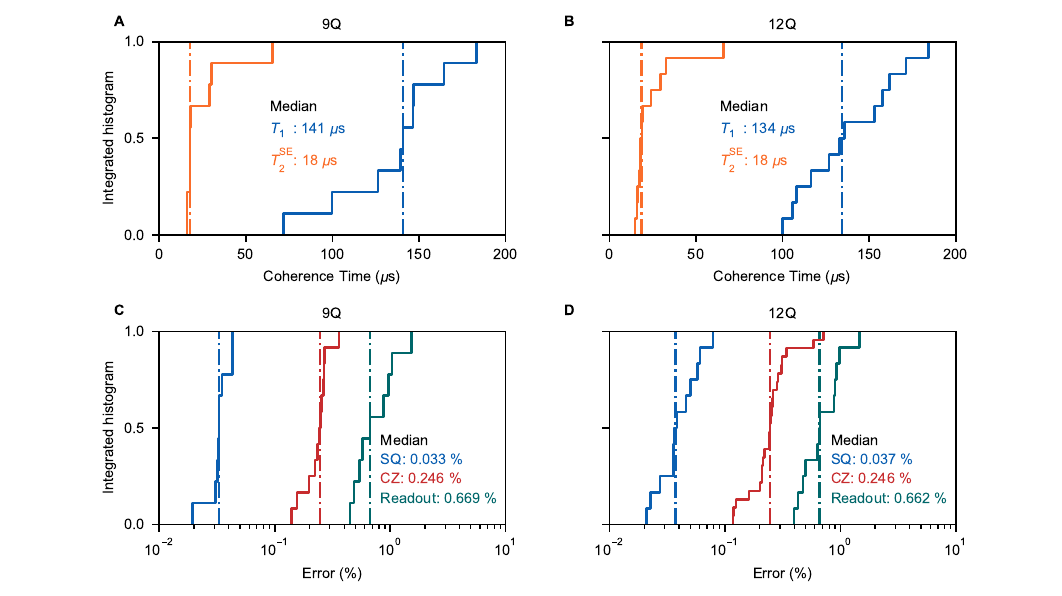}
    \end{center}
    \caption{{\bf Integrated histograms of typical qubit performance for 9-qubit and 12-qubit systems.}  
    ({\bf A-B}) Statistics of qubit energy relaxation time $T_1$ (blue lines) and spin-echo dephasing time $T_2^{\rm{SE}}$ (orange lines) for 9-qubit and 12-qubit systems. 
    ({\bf C-D}), Statistics of single-qubit gate error (SQ, blue lines), two-qubit CZ gate error (CZ, red lines), and readout error (green lines) for 9-qubit and 12-qubit systems. Gate errors are obtained by simultaneous cross-entropy benchmarking (XEB)~\cite{Boixo2018NP}. The readout error is calculated as the mean value of $|0\rangle$ and $|1\rangle$ readout errors.
    The dashed-dotted lines indicate the median values of these parameters.}
    \label{smfig:device_information}
\end{figure}

Our experiments are performed on a superconducting quantum processor possessing 121 frequency-tunable transmon qubits and 220 tunable couplers~\cite{xu2023digital}, where we choose a $3\times 3$ and a $3\times 4$ two-dimensional sub-lattice to explore the robustness of quantum Darwinism (Fig.~3 of the main text) and the emergent branching structure (Fig.~2 of the main text), respectively. The typical performance for these qubits is shown in Fig.~\ref{smfig:device_information}. The median values of energy relaxation time $T_1$ and spin-echo dephasing time $T_2^{\rm SE}$ for 9-qubit (12-qubit) system are 141 $\mu$s (134 $\mu$s), and 18 $\mu$s (18 $\mu$s) respectively, as depicted in Fig.~\ref{smfig:device_information}A and B. For the 9-qubit (12-qubit) configuration, the median Pauli errors of single-qubit gate, two-qubit CZ gate, along with the median readout error are 0.033\% (0.037\%), 0.246\% (0.246\%) and 0.669\% (0.662\%), which are shown in Fig.~\ref{smfig:device_information}C and D. See Ref.~\cite{xu2023digital,Bao2024NC} for further details of the experimental setup. 

\subsection{Quantum state tomography\label{sec:tomo}}
To measure the information-theoretical quantities, such as mutual information and quantum discord, we need the density matrices of the system $\cS$ and environment $\cE$, which can be obtained by performing quantum state tomography (QST)~\cite{1989_Vogel_PRA}. To this end, we apply $\{I,~R_x(\pi/2),~R_y(\pi/2)\}^{\otimes n}$ gates to the related qubits to rotate the measurement basis; this means that one needs to repeat the circuit $3^n$ times, where $n$ is the number of qubits. Therefore, in Figs.~2B and 3B of the main text, the circuits end up with tomography gates in $\{I,~R_x(\pi/2),~R_y(\pi/2)\}^{\otimes n}$ before measurements. With these measurement results, an over-determined equation can be solved to reconstruct the density matrix. In practice, the qubit number $n$ varies from 1 to 5. To reduce the measurement time, we only measure $n=5$ and trace out some qubits to get density matrices for $n<5$. Note that QST is unscalable for experiments since an exponentially increasing number of tomography operations are needed.

\subsection{Measurement of the geometric state}
In the main text, we have shown that any composite state of the system $\cS\cE$ can be written as
\begin{align}
\left|\psi_{\mathcal{SE}}\right\rangle = \sum_{\alpha, \beta} \sqrt{X_{\alpha \beta}}\left|\chi_{\alpha\beta}\right\rangle\left|f_\alpha\right\rangle\left|\bar{f}_\beta\right\rangle.
\end{align}
For the 12-qubit lattice, we consider the two central qubits as the system, whose pointer states are described by logical states $|0_{\cS}\rangle = |0_{\rm L}\rangle=|00\rangle$ and $|1_{\cS}\rangle=|1_{\rm L}\rangle=|11\rangle$. Each environment basis $|f_{\alpha}\rangle|\bar{f}_{\beta}\rangle$ encodes a system state $|\chi_{\alpha\beta}\rangle=\sqrt{\frac{X_{0,\alpha\beta}}{X_{\alpha\beta}}}|0_{\rm L}\rangle+\sqrt{\frac{X_{1,\alpha\beta}}{X_{\alpha\beta}}}|1_{\rm L}\rangle$ with a probability $X_{\alpha\beta}$. To obtain the system state $|\chi_{\alpha\beta}\rangle$ and the corresponding environment basis $|f_{\alpha}\rangle|\bar{f}_{\beta}\rangle$, we apply logical QST gates
\begin{align}
    I_{\rm L} &= I\otimes I,\\
    Rx_{\rm L}(\pi/2) &= \cos(\pi/4)I\otimes I-i\sin(\pi/4)X\otimes X,\\
    Ry_{\rm L}(\pi/2) &=\cos(\pi/4)I\otimes I-i\sin(\pi/4)Y\otimes Y,
\end{align}
to the central qubits and simultaneously record the measurement outcomes of the environment basis $|f_{\alpha}\rangle|\bar{f}_{\beta}\rangle$. In our experiments, we only keep the measurement outcomes that are in the logical space by post-selection, while those results that leak out of the logical basis due to experimental imperfections, e.g., gate errors, are discarded. Then the experimental $\rho_{\alpha\beta}$ of $|\psi_{\alpha\beta}\rangle$ can be extracted using the method mentioned in Section \ref{sec:exp}B. The state $\rho_{\alpha\beta}$ can then be visualized on the Bloch sphere with coordinates calculated by $\left({\rm Tr}(\rho_{\alpha\beta}\sigma_x),~{\rm Tr}(\rho_{\alpha\beta}\sigma_y),~{\rm Tr}(\rho_{\alpha\beta}\sigma_z)\right)$. In the experimental realization, $Rx_{\rm L}(\pi/2)$ and $Ry_{\rm L}(\pi/2)$ are further decomposed into single-qubit rotations and controlled-Z (CZ) gates, where the matrix form of CZ gate is  
\begin{align}
    U_{\rm CZ}=\begin{bmatrix}
        1&0&0&0\\
        0&1&0&0\\
        0&0&1&0\\
        0&0&0&-1\\
    \end{bmatrix}.
\end{align}

\subsection{Measurement of the quantum discord}
In Fig. 3 of the main text, we obtain the density matrix of the central qubit and four environment qubits $\rho_{\mathcal{S}\mathcal{F}}$ by QST. Here, we introduce how we calculate the quantum discord with $\rho_{\mathcal{S}\mathcal{F}}$. Quantum discord is defined as~\cite{Zurek2000AP,Ollivier2001PRL}
\begin{align}
    \discord=\MI-\holevo.
\end{align}
We can obtain mutual information $\MI$ straightforwardly for $m=1,~2,~3$ and 4 by calculating the von Neumann entropy of the system and environment,
\begin{align}
    \MI=H_{\mathcal{S}}+H_{\mathcal{F}}-H_{\mathcal{S}\mathcal{F}}.
\end{align}
However, the Holevo bound $\holevo$ requires calculating the conditional entropy 
\begin{align}
    H_{\mathcal{S}|\mathcal{\check{F}}} = \sum_{\alpha} p_{\alpha}H_{\mathcal{S}||f_{\alpha}\rangle},
\end{align}
where $p_{\alpha}=\Tr{[\rho_{\mathcal{S}\mathcal{F}} (I\otimes|f_\alpha\rangle\langle f_{\alpha}|)]}$, is the probability that $\mathcal{F}$ is at state $|f_{\alpha}\rangle$.  Then we can calculate the conditional density matrix as
\begin{align}
    \rho_{\mathcal{S}||f_{\alpha}\rangle}={\rm Tr}_{\mathcal{F}}\left(\frac{M_{\alpha} \rho_{\mathcal{S}\mathcal{F}} M_{\alpha}^{\dagger}}{{\rm Tr}\left(M_{\alpha} \rho_{\mathcal{S}\mathcal{F}} M_{\alpha}^{\dagger}\right)}\right),
\end{align}
where $M_{\alpha}=|f_{\alpha}\rangle\langle f_{\alpha}|$.

\subsection{Optimization of experimental circuit }
\begin{figure}[h!]
    \begin{center}
    \includegraphics[width=1\textwidth]{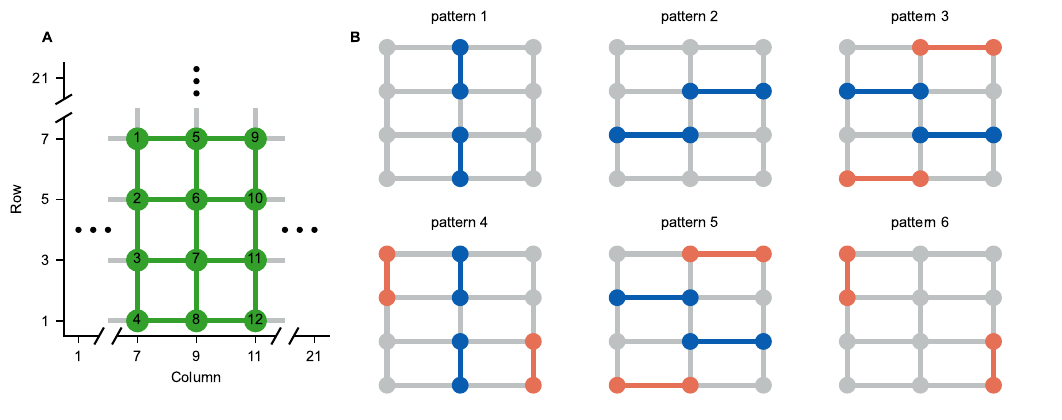}
    \end{center}
    \caption{{\bf The chip layout and two-qubit gate patterns used for observing emergent classical reality.} ({\bf A}) The physical layout for the 12-qubit system. The circles denote qubits, and the lines represent couplers. The system consists of two qubits living in the center of the lattice (Q${}_6$ and Q${}_7$), surrounded by ten environment qubits. ({\bf B}) Six two-qubit gate patterns corresponding to the circuit shown in Fig.~2B in the main text (the $U_{\rm tomo.}$ is not displayed here). In each two-qubit gate pattern, two blue circles connected by a blue line depict the interaction between the system and the environment. Orange circles and lines represent SWAP gates. In the experiment, all two-qubit gates are decomposed into single-qubit rotations and CZ gates.}
    \label{smfig:patterns}
\end{figure}
The circuit shown in Fig.~2B of the main text is a conceptual representation of the 12-qubit model in Fig.~2A of the main text, which requires long-range connectivity between the qubits. However, our physical device is a rectangular lattice (Fig.~\ref{smfig:patterns}A) with nearest-neighboring coupling, which cannot realize all the interactions between the system and environment directly. For example, as depicted in Fig.~\ref{smfig:patterns}B, the four corner qubits cannot interact with the central qubits directly. To overcome this, we utilize SWAP gates (orange lines) to bridges the interaction between the corner qubits and the system qubits. In our experiment, it is convenient to realize high-fidelity arbitrary single-qubit gates and CZ gates instead of random $U_k$ conditional gates or SWAP gates. On the other hand, reducing the circuit depth and the number of gates is crucial to mitigate circuit errors. Therefore, we use Qiskit~\cite{qiskit2024} to transpile the raw circuits into the desired ones which consist of single-qubit rotations and CZ gates.

\subsection{Noisy simulation}
\begin{table}[h!]
    \centering
    \caption{The parameters and error rates used in the noisy simulation.}
    \begin{tabular}{|c|c|c|}
        \hline \diagbox{Parameters (mean)}{System}&9-qubit&12-qubit \\
        \hline
        Single-qubit gate time $t_{\rm SQ}$ & 20~ns& 20~ns\\
        \hline
        Single-qubit idle gate time $t_{\rm SQ,idle}$ & 47~ns& 47~ns\\
        \hline
        CZ gate time $t_{\rm CZ}$ & 47~ns& 47~ns\\
        \hline
        Single-qubit gate error $\epsilon_{\rm SQ}$ & 0.033~\% & 0.043~\% \\
        \hline
        Single-qubit idle gate error $\epsilon_{\rm SQ,idle}$ & 0.082~\% & 0.079~\% \\
        \hline
        CZ gate error $\epsilon_{\rm CZ}$ & 0.238~\% & 0.277~\% \\
        \hline
        Readout error $\epsilon_{\rm readout}$ & 0.8~\% & 0.8~\%\\
        \hline
        $T_1$ & 135~$\mu$s& 135~$\mu$s\\
        \hline
        $T_{\phi}$ & 40~$\mu$s& 40~$\mu$s\\
        \hline
        shots & 5,000,000& 1,000,000\\
        \hline
    \end{tabular}
    \label{tab:noisy_sim}
\end{table}
In the main text, we perform the noisy simulation following the procedure described in Ref.~\cite{xiang2024nc}. There are various experimental errors, among which we mainly consider decoherence, depolarizing, and readout error (see Table~\ref{tab:noisy_sim}). More specifically, we use Qiskit's~\cite{qiskit2024} noisy simulator to randomly sample the corresponding error operators and perform this process under different measurement bases for tomography. This process is similar to what is carried out experimentally. The related parameters and error rates listed in Table~\ref{tab:noisy_sim} are also consistent with the experimental measurements. Note that $\epsilon_{\rm SQ,idle}$ is applied to those idle qubits when executing CZ gates in the same layer.

\bibliography{sm}
